\documentclass[twocolumn,prl]{revtex4}
\usepackage{graphicx}
\usepackage{amsmath}

\begin{document}
\title{Distribution of time-bin entangled qubits over 50\,km of optical fiber}
\author{I. Marcikic, H. de Riedmatten, W. Tittel, H. Zbinden, M. Legr\'{e} and N.
Gisin}

\affiliation{Group of Applied Physics-Optique, University of
Geneva, CH-1211, Geneva 4, Switzerland}

\begin{abstract}

We report experimental distribution of time-bin entangled qubits
over 50\,km of optical fibers. Using actively stabilized
preparation and measurement devices we demonstrate violation of
the CHSH Bell inequality by more than 15 standard deviations
without removing the detector noise. In
addition we report a proof of principle experiment of quantum key distribution over 50\,km of optical
fibers using entangled photon.

\end{abstract}

\maketitle

In the science of quantum information a central experimental issue
is how to distribute entangled states over large distances.
Indeed, most protocols in quantum communication require the
different parties to share entanglement. The best-known examples
are Quantum Teleportation \cite{brassard93} and Ekert's Quantum
Key Distribution (QKD) protocol \cite{ekert}. Note that even in
protocols that do not explicitly require entanglement, like the
BB84 QKD protocol \cite{bb84}, security proofs are often based on
"virtual entanglement", i.e. on the fact that an ideal single
photon source is indistinguishable from an entangled photon pair
source in which one photon is used as a trigger
\cite{ShorPreskill00}. From a more practical point of view,
entanglement over significant distances can be used to increase
the maximal distance a quantum state can cover, as in quantum
repeater \cite{briegel98} and quantum relay \cite{relay}
protocols. Finally, entanglement is also treated as a resource in
the study of communication complexity \cite{Brassard03}.

As entanglement cannot be created by shared randomness and local
operations, it must be somehow distributed. Recently there have
been some proposals to use satellites for long distance
transmission \cite{aspelmeyer03}. Also some experiments through
open space have been performed either for QKD (over 50\,m)
\cite{beveratos02} or for the transmission of entangled qubits
(over 600\,m) \cite{aspelmeyer031}. Despite the weather and
daylight problems, this is an interesting approach. Another
possibility, that we follow in this work, is to use the worldwide
implemented optical fiber network. This, however, implies some
constraints. One should operate at telecommunication wavelengths
(1.3 or 1.55\,$\mu$m), in order to minimize losses in optical
fibers, and the encoding of the qubits must be robust against
decoherence in optical fibers. Likely the most adequate way to
encode qubits is to use energy-time \cite{franson89} or it's
discrete version time-bin encoding \cite{jurgen99}. The major
drawback of this kind of encoding, compared to polarization type,
is that the creation and the measurement is more complex: it
relies on stable interferometers. In this letter we report a way
to create and to measure time-bin entangled qubits which allows us
to violate Bell inequalities over 50\,km of optical fibers and to
show a proof of principle for entanglement based QKD over long
ranges. Moreover it allows to demonstrate stability of our entire
set-up over several hours.

\begin{figure}[h]
\includegraphics[width=7cm]{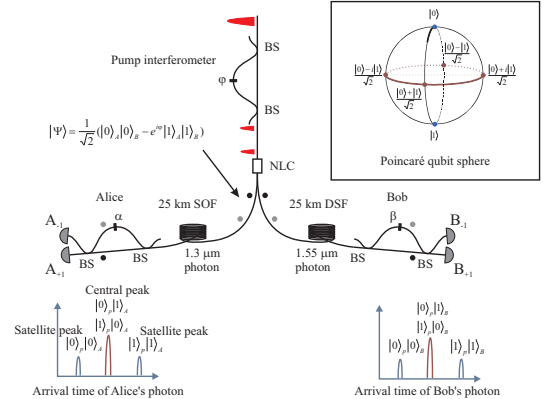}
\caption{{Scheme of the experimental set-up. Time bin qubits are
prepared by passing a fs pulse through the pump interferometer and
a non-linear crystal (NLC). Eventually, a pair of entangled
photons is created in the crystal. They are sent to Alice and Bob
through 25.3\,km of optical fibers. Alice and Bob analyze photons
using interferometers equally unbalanced with respect to the pump
interferometer. All three interferometers are built using passive
50-50 beam-splitters (BS). Alice's and Bob's detection times are
also represented. }} \label{setup}
\end{figure}

Let us first remind the reader how to create and measure time-bin
entangled qubits. They are created by sending a short laser pulse
first through an unbalanced interferometer (denoted as the pump
interferometer) and then through a non-linear crystal where
eventually a pair of photons is created by spontaneous parametric
down conversion (SPDC)(see Fig.\ref{setup}). The state can be
written:
\begin{equation}
\left| \Psi \right\rangle =\frac{1}{\sqrt{2}}(\left|
0\right\rangle _{A}\left| 0\right\rangle _{B}-e^{i\varphi}\left|
1\right\rangle _{A}\left|1\right\rangle _{B}) \label{2}
\end{equation}
where $\left|0\right \rangle$ represents a photon in the first
time bin (having passed through the short arm) and $\left|1\right
\rangle$ a photon in the second time-bin (having passed through
the long arm). The index $A$ and $B$ represents Alice's and Bob's
photon. The phase $\varphi$ is defined with respect to a reference
path length difference between the short and the long arm $\Delta
\tau$.
The photons A and B are then sent to Alice and Bob who perform
projective measurements, by using a similar unbalanced
interferometer. There are three detection times on Alice's (Bob's)
detectors with the respect to the emission time of the pump laser
(see Fig.\ref{setup}). The first and the last peak (denoted as
satellite peaks) corresponds to events which are temporally
distinguishable: the left (right) peak corresponds to a photon
created in the first (second) time-bin which passed through the
short (long) arm of Alice's interferometer. When detected in the
left (right) satellite peak, the photon is projected onto the
vector $\left|0\right \rangle$ ($\left|1\right \rangle$) (the
poles on the Poincar\'e qubit sphere). Photons detected in the
central peak can be either due to events where the created photon
is in the first time-bin and then it passes through the long arm
of Alice's interferometer or due to events where the photon is
created in the second time-bin and then passes through the short
arm of Alice's interferometer. In this case the photon is
projected onto the  vector $\left|0\right\rangle
+e^{i\alpha}\left|1\right\rangle$ (i.e. on the equator of the
Poincar\'e qubit sphere). Note that when Alice records the central
peak she does not observe single photon interference by changing
the phase of her interferometer because which-path information can
be found by recording the emission time of Bob's photon. With
reference to experiments using polarization entangled photons, we
refer to this as rotational invariance \cite{clauser74}. If Alice
and Bob both record counts in their central peaks, they observe
second order interference by changing either the phase in Alice's,
in Bob's or in the pump interferometer. The coincidence count rate
between Alice's and Bob's detectors $A_iB_j$, is then given by:
\begin{equation}
R_{A_i,B_j}(\alpha,\beta,\varphi)\sim
1+ijVcos(\alpha+\beta-\varphi) \label{3}
\end{equation}
where $i$ and $j=\pm 1$ (see Fig.\ref{setup}) and V is visibility
of the interference fringes (which can in principle reach the
value of 1). We define the imbalance of the pump interferometer as
the reference time difference $\Delta\tau$ between the first and
the second time-bin, the phase $\varphi$ is thus taken to be zero.
The correlation coefficient is defined as:
\begin{equation}
E(\alpha,\beta)=\frac{\displaystyle \sum_{i,j}
ijR_{A_iB_j}(\alpha,\beta)}{\displaystyle \sum_{i,j}
R_{A_iB_j}(\alpha,\beta)} \label{4}
\end{equation}
and by inserting Eq.\ref{3} into Eq.\ref{4} the correlation
coefficient becomes:
\begin{equation}
E(\alpha,\beta)=Vcos(\alpha+\beta) \label{5}
\end{equation}

The Bell inequalities define an upper bound for correlations that
can be described by local hidden variable theories (LHVT). One of
the most frequently used forms, known as the
Clauser-Horne-Shimony-Holt (CHSH) Bell inequality \cite{clauser},
is:
\begin{equation}
S=\vert
E(\alpha,\beta)+E(\alpha,\beta')+E(\alpha',\beta)-E(\alpha',\beta')
\vert \leq 2 \label{6}
\end{equation}
Quantum mechanics predicts that $S$ has a maximum value of
$S=2\sqrt{2}$ with $\alpha= 0^{\circ}$,$\alpha'= 90^{\circ}$,
$\beta=45^{\circ}$ and $\beta'=-45^{\circ}$. It has been also
shown that when the correlation function has sinusoidal form of
Eq.\ref{5} and when there is rotational invariance, the boundary
condition of Eq.\ref{6} can be written as:
\begin{equation}
S=2\sqrt{2}V \leq 2\label{7}
\end{equation}
thus $V \geq \frac{1}{\sqrt{2}}$ implies violation of the CHSH
Bell inequality, i.e. correlations can not be explained by LHVT.

Our experimental set-up is the following (see Fig.\ref{setup}): A
150\,femtosecond laser pulse with a 710\,nm wavelength  and with a
repetition rate of 75\,MHz is first sent through an unbalanced,
bulk, Michelson interferometer with an optical path difference of
$\Delta\tau=1.2$\,ns and then through a type I LBO (lithium
triborate) non-linear crystal where collinear non-degenerate
photon pairs at 1.3 and 1.55\,$\mu$m wavelength can be created by
SPDC. The pump beam is then removed with a silicon filter and the
pairs are coupled into an optical fiber. The photons are separated
with a wavelength-division-multiplexer, the 1.3\,$\mu$m photon is
sent through 25.3\,km of standard optical fiber (SOF) to Alice and
the 1.55\,$\mu$m photon through 25.3\,km of dispersion shifted
fiber (DSF) to Bob \cite{explication1}. Alice's photon is then
measured with a fiber Michelson interferometer and detected by one
of two liquid nitrogen cooled passively quenched Germanium
avalanche photo-diodes (APD) $A_{+1}$ or $A_{\text{\small{-}}1}$.
Their quantum efficiency is of around 10\,\% with 20\,kHz of dark
counts. In order to select only the central peak events and also
to reduce the detector dark counts, a coincidence is made with the
emission time of the laser pulse. This signal then triggers Bob's
detectors ($B_{+1}$ and $B_{\text{\small{-}}1}$) which are two
InGaAs APDs (IdQuantique) working in so called gated mode.
Although both detectors have similar quantum efficiencies of
20\,\%, one of the detectors ($B_{+1}$) dark count probability is
two times smaller than the other one ($B_{\text{\small{-}}1}$),
and is around $10^{-4}$\,/ns. To reduce chromatic dispersion in
optical fibres and the detection of multiple pairs
\cite{marcikic02}, we use interference filters with spectral width
of 10\,nm for 1.3\,$\mu$m photons and 18\,nm for the 1.55\,$\mu$m
photons. Using 70\,mW of average input power (measured after the
pump interferometer) the probability of creating an entangled
qubit per pulse is around 8\,\%. Bob's analyzer is also a
Michelson type interferometers built with optical fibers. To
better control the phase and to achieve long term stability all
three interferometers are passively and actively stabilized.
Passive stabilization consists of controlling the temperature of
each interferometer. Active stabilization consists of probing the
interferometer's phase with a frequency stabilized laser at
1.534\,$\mu$m (Dicos), and to lock them to a desired value via a
feedback loop on a piezo actuator (PZA) included in each
interferometer. In order to change path difference in the bulk
pump-interferometer, one of the mirrors is mounted on a
translation stage including a PZA with the range of around
4\,$\mu$m. In the analyzing interferometers the long fiber path is
wound around a cylindric PZA with a circumference variation range
of 60\,$\mu$m. Contrary to the bulk interferometer which is
continuously stabilized, the phase of the fiber interferometers
can not be stabilized during the measurement period. Thus we
continuously alternate between measurement periods of 100\,seconds
and stabilization periods of 5\,seconds. This method allows us not
only to stabilize the entire set-up during several hours, but also
to have good control over the changes of both phases $\alpha$ and
$\beta$.

\begin{figure}[h]
\includegraphics[width=8.43cm]{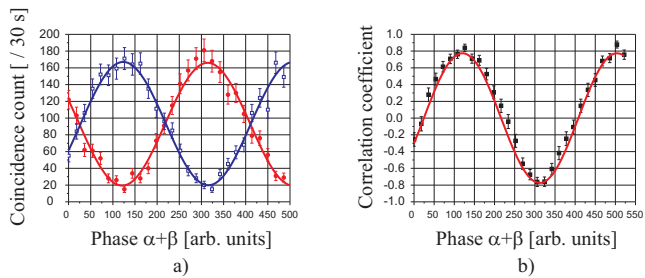}
\caption{{a) Coincidence counts between detectors $A_{+1}B_{+1}$
(circles) and $A_{+1}B_{\text{\small{-}}1}$ (open squares) b)
Correlation coefficient E($\alpha,\beta$) measured from four
different coincidence counts (Eq.\ref{4}). Alice's phase $\alpha$
is kept constant and Bob's phase $\beta$ is scanned}} \label{corr}
\end{figure}

In order to show a violation of the CHSH Bell inequality after
50\,km of optical fibers, we proceed in two steps: first we scan
Bob's phase $\beta$ while Alice's phase $\alpha$ is kept constant.
We obtain a raw visibility of around $78\pm1.6$\,\% (see
Fig.\ref{corr}) from which we can infer an $S$ parameter of
$S=2.206\pm0.045$ (Eq.\ref{7}) leading to a violation of the CHSH
Bell inequality by more than 4 standard deviations. The
coincidence count rate between any combination of detectors
$A_iB_j$ is of around 3\,Hz.
The raw visibility of the correlation function is mainly reduced
due to the creation of multiple pairs (around 9\,\%), due to
accidental coincidence counts (related to dark counts of our
detectors, around 8\,\%) and due to the misalignment of the
interferometers (around 5\,\%). In principle one could reduce the
creation of multiple pairs by reducing the input power, but then
the coincidence count rate would also decrease.

\begin{figure}[h]
\includegraphics[width=8.43cm]{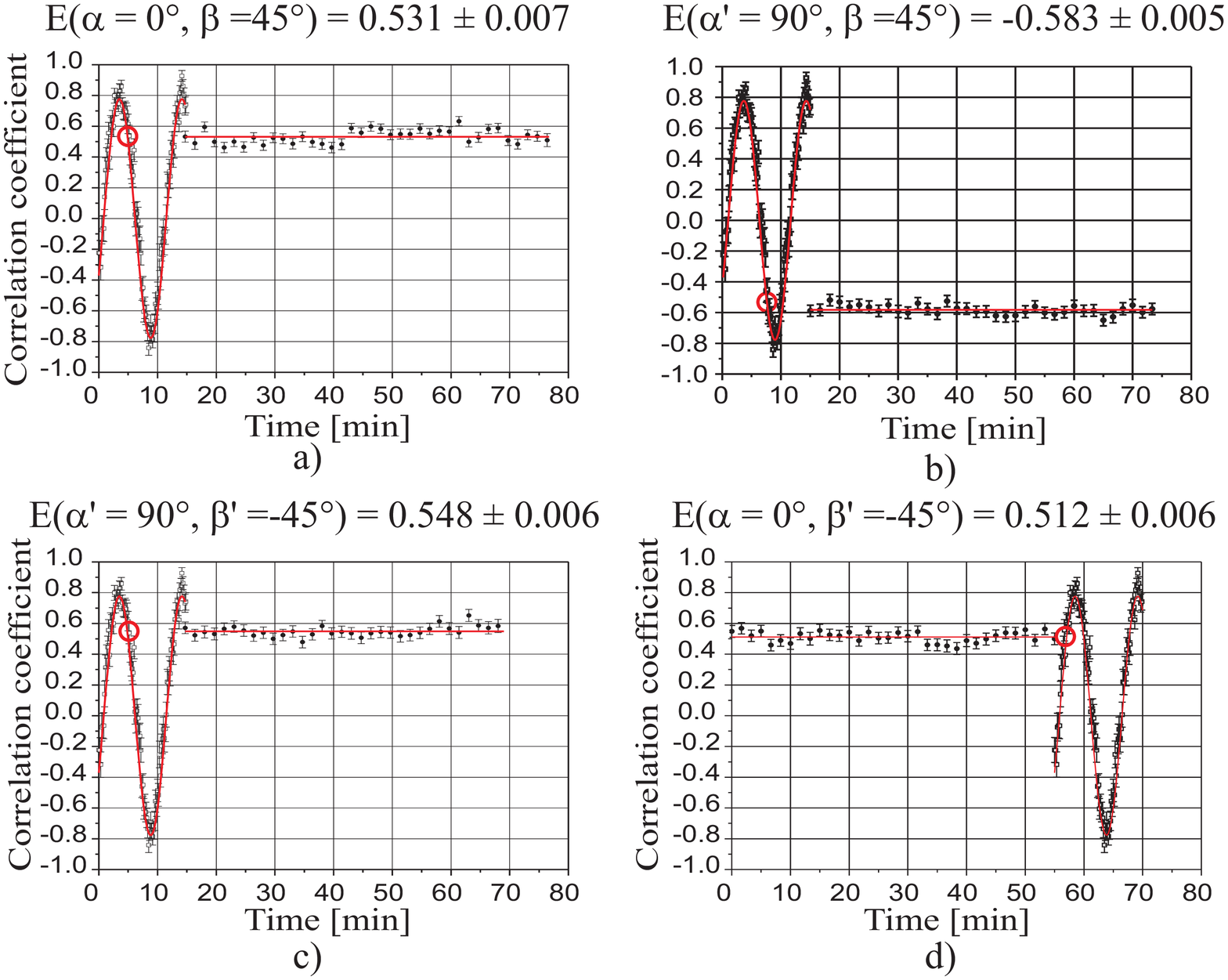}
\caption{{Correlation coefficients for continuous scan and four
different settings. Each data point is derived from a 100\,s
integration time of coincidence counts between four different
combinations of two detectors (Eq.\ref{4}). As $\alpha$ and
$\beta$ are defined relatively to the pump-interferometer's phase,
we use the first three measurement a), b) and c) to define four
different phases: $\alpha=0^{\circ}$, $\alpha'=90^{\circ}$,
$\beta=45^{\circ}$ and $\beta'=-45^{\circ}$. The last measurement
d) completes the proof of a violation of the CHSH Bell inequality.
The open circles represent the correlation coefficient value for
which the CHSH Bell inequality would be maximally violated when
the maximum visibility is 78\,\%.}} \label{fset}
\end{figure}

With our new interferometers we are able to perform for the first
time with time-bins the second step: measure the CHSH Bell
inequality according to Eq.\ref{6}, i.e. lock the phase to the
desired value in order to measure the four different correlation
coefficients one after the other. To reduce statistical
fluctuations, we measure the correlation coefficient (Eq.\ref{4})
during almost an hour for each setting. The obtained $S$ parameter
is $S=2.185\pm0.006$ which shows a violation of the CHSH Bell
inequality by more than 15 standard deviations (see
Fig.\ref{fset}).

It has been proven that when the Bell inequality is violated the
entangled photons can be used in quantum cryptography \cite{rmp}.
Our QKD protocol is analogous to the BB84 protocol using time-bin
entangled photons \cite{tittel00}. Hence, Alice and Bob use two
maximally conjugated measurement basis. The first basis is defined
by two orthogonal vectors $\left| 0 \right \rangle$ and $\left|1
\right \rangle$ represented on the poles of the Poincar\'{e} qubit
sphere (Fig.\ref{setup}). The projection onto this basis is
performed whenever a  photon is detected in a satellite peak. Let
us illustrate how Alice and Bob encode their bits: whenever Alice
detects her photon in the first (second) satellite peak she knows
that the pair is created in the first (second) time-bin and thus
Bob can either detect the twin photon in the first (second)
satellite peak or in the central peak, however he can never detect
it in the second (first) satellite peak. Thus, after suppressing
central peak events with the basis reconciliation, Alice and Bob
encode their bits as 0 (1) if the photon is detected in the first
(second) satellite peak. The second basis is defined by two
orthogonal vectors represented on the equator of the Poincar\'{e}
sphere (for example $\frac{\left| 0 \right \rangle+\left| 1 \right
\rangle}{\sqrt{2}}$ and $\frac{\left| 0 \right \rangle-\left| 1
\right \rangle}{\sqrt{2}}$). The projection onto this basis is
performed when a photon is detected in the central peak. Alice and
Bob have to correctly adjust their interferometers such that they
have perfect correlation between detectors $A_{+1}B_{+1}$ and
$A_{\text{\small{-}}1}B_{\text{\small{-}}1}$. The encoding of bits
0 and 1 in this basis is thus defined by which detector fires. As
Alice's and Bob's photon passively choose their respective
measurement basis, there is 50\,\% probability that they are
detected in the same basis which ensures the security against
photon number splitting attack \cite{rmp}.

\begin{figure}[h]
\includegraphics[width=8.43cm]{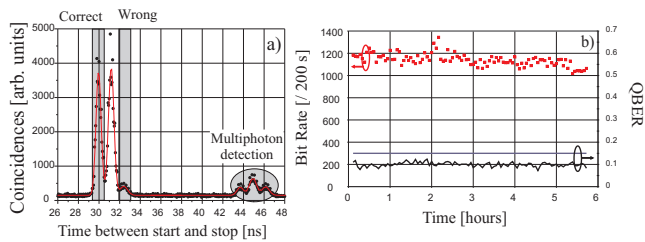}
\caption{{Experimental results. a) Coincidence count between
Alice's and Bob's detector where Alice selects bit 0 in the first
measurement basis. Bob detects photons projected onto
$\left|0\right\rangle$ vector (denoted as correct events) or onto
$\left|0\right\rangle+e^{i\beta}\left|1\right\rangle$ vector
(these events are removed by basis reconciliation).  The presence
of multiphotons leads to wrong detections and thus to the increase
of the QBER. b) Bit rate results for the second basis (squares)
and a QBER measurement (line), which is clearly below the QBER
limit of 15\,\% secure against individual attacks (straight line)
\cite{Fuchs97}.}} \label{crypto}
\end{figure}

We report a proof of principle of entanglement based QKD over
50\,km of optical fiber. In our experimental set-up, Alice
sequentially selects one of the three detection windows by looking
at the arrival time of her photon with respect to the emission of
the laser pulse (see Fig.\ref{setup}). This signal is then used to
trigger Bob's detectors. In the first measurement basis the
measured quantum bit error rate (QBER) \cite{explication2} is of
$12.8\pm0.1$\,\% and the measured raw bit rate of around 5\,Hz.
The QBER is due to accidental coincidence counts (around 8\,\%)
and to creation of multiple pairs (around 4.5\,\%, see
Fig.\ref{crypto}a)). In the second measurement basis the measured
QBER is of $10.5\pm0.09$\,\% (Fig.\ref{crypto}b)), with a bit rate
of 6\,Hz. In this case the QBER is due to accidental coincidence
count probability (around 4\,\%), to creation of multiple pairs
(around 4.5\,\%) and to slight misalignment of our interferometers
(around 2\,\%). In order to have a low statistical error the
integration time for both basis is of around six hours. The
difference of the QBER measured in two basis is due to the fact
that in the first measurement basis the detectors are opened
during  two time-windows instead of one in the second basis.
However in the first basis the misalignment of interferometers
does not introduce any error. Note that by using two InGaAs APDs
with the same low dark count probability as detector $B_{+1}$, the
QBER in the first measurement basis would be reduced to 10.8\,\%
and in the second basis to 9.8\,\%.

For a true implementation of QKD using time-bin entangled photons
it is necessary that Alice and Bob can monitor detections in all
three time windows at the same time and not as presented here, one
after the other. In addition, as Alice has to trigger Bob's
detectors, it is important to ensure that Eve does not get any
information about Alice's detection times. This extensions would
require more coincidence electronics but can be easily
implemented. Finally, note that Alice's trigger signal has to
arrive at Bob's before the photon, thereby putting constraints on
the distance between Alice, Bob and the source of entangled
photons. These limitations are suppressed by using passively
quenched InGaAs APDs (work in progress) or detectors based on
superconductivity \cite{sobolewski03}.

In this letter we present an experimental distribution of time-bin
entangled photons over 50\,km of optical fiber. Using active phase
stabilization with a frequency stabilized laser and feedback loop,
long term stability and control of the interferometer's phase is
achieved. In the first experiment, the CHSH Bell inequality is
violated by more than 15 standard deviation without removing the
detector noise. The possibility of changing the phase in a
controlled way allowed us also to show a proof of principle of
entanglement based quantum key distribution over 50\,km of optical
fiber. An average Quantum Bit Error Rate of 11.5\,\% is
demonstrated which is small enough to establish quantum keys
secure against individual attacks \cite{Fuchs97}. Finally, a long
term set-up stability opens the road for future demonstrations of
more complicated quantum communication protocols requiring long
measurement times as is the case for the entanglement swapping
protocol.

The authors would like to thank Claudio Barreiro and Jean-Daniel
Gautier for technical support. Financial support by the Swiss NCCR
Quantum Photonics, and by the European project RamboQ are
acknowledged.


\begin{thebibliography}{99}
\bibitem{explication2} The QBER is defined as the ratio of
error rate to total rate.
\end{thebibliography}
\end{document}